# Electrical measurement of antivortex wall eigenfrequency

Mahdi Jamali, Jae Hyun Kwon, Kulothungasagaran Narayanapillai, and Hyunsoo Yang[*]

*Department of Electrical and Computer Engineering, National University of Singapore, 4 Engineering Drive 3, Singapore 117576, Singapore*

The dynamics of a ferromagnetic antivortex wall has been experimentally studied in a magnetic nanostructure. Two different techniques have been used to independently measure the eigenfrequency of an antivortex wall such as the resonance excitation by sinusoidal microwave and the damped resonance excitation induced by short voltage pulses. Direct observation of antivortex wall nucleation has been measured in the frequency domain for the first time. Electrical measurements of the antivortex dynamics in frequency domain reveal the existence of multi-eigenmodes as well as nonlinear behaviors for large excitation amplitudes. The time resolved measurements of the antivortex wall show that the frequency of the damped gyration is similar to that of frequency domain and coexistence of spin wave excitations.

The determination of eigenfrequencies and eigenmodes, which mainly depend on the physical properties of a system such as materials, shapes, and dimensions, of nanostructures are crucial for the proper understanding of their dynamics under different excitations. Recently, there have been substantial efforts in understanding dynamics of vortex-like structures from magnetic nanostructures due to their potential in memory applications [1, 2]. There are two types of vortex-like structures in a quasi two dimensional ferromagnet; one is a circular structure called the vortex wall, and the other one is an antivortex. Both structures contain a core which is perpendicular to the plane of magnetization. There have been numerous reports on the measurements of the vortex wall dynamics by optical techniques [3-5] and electrical methods [6-8]. The antivortex wall exhibits promising properties such as the ultrafast switching of the core [9-12]. However, stable antivortices have only been observed in a few types of ferromagnetic structures [13, 14]. While there have been some experimental reports on the measurement of the antivortex using optical methods [15, 16] in the past few years, to the best of our knowledge, there has been no study on the electrical measurement of the antivortex dynamics. We present the first all-electrical measurement of antivortex dynamics. Although optical techniques give more insight about the details of the antivortex dynamics such as the domain wall trajectories [15], the all-electrical measurement is an important step toward realization of the antivortex based devices for applications.

In this Letter, we have used two different techniques to measure the dynamical properties of the antivortex, including eigenmode frequencies and damping factor. From our measurements, it is found that by using the domain wall rectification effect [6, 17, 18], it is possible to detect the nucleation of the antivortex wall in the presence of a perpendicular magnetic field. Multiple eigenmodes have been observed in the frequency spectrum of the antivortex wall. Furthermore, the resonance frequency of the antivortex wall depends weakly on the in-plane external magnetic field at low bias fields, and we observed a sizable shift in the resonance frequency of the antivortex at high in-plane magnetic fields. On increasing the excitation amplitude, higher frequency modes have been detected. In addition, decreasing of the device dimension results in an increase in the resonance frequency of the antivortex. The time-resolved measurement of the antivortex wall response to a narrow voltage pulse (< 10 ns) has been performed. The antivortex frequency of the damped gyration is found to be similar to the resonance frequency of the antivortex, measured by the antivortex rectification effect. Furthermore, the excitation of antivortex dynamics coexists with spin wave excitations inside the ferromagnetic structure.

Antivortex dynamics has been studied on the infinity-shaped ferromagnetic nanostructure shown in Fig. 1(a). Devices are fabricated, first by the dc sputter deposition of Ta (2 nm)/Ni$_{81}$Fe$_{19}$ (20 nm )/Ta (2 nm ) on a Si/SiO$_2$ (300 nm) substrate, followed by the patterning of the infinity-shaped nanostructure using e-beam lithography and argon ion milling. 2 nm SiO$_2$ is rf sputter deposited after ion milling without breaking vacuum to avoid the oxidation of the ferromagnetic nanostructure. A second e-beam lithography step is used to pattern Ta (5 nm)/Cu (100 nm) contacts. Before contact deposition, the interface between the contacts and the ferromagnetic structure was etched through 2 nm to clean the interface. The SEM image of the device structure is shown in Fig. 1(a).

In order to measure domain wall dynamics, we have used the domain wall rectification effect, and the domain wall resonance frequency has been measured using a homodyne circuit shown in Fig. 1(a). A low frequency voltage from a lock-in amplifier is mixed with a high frequency sinusoidal signal from a microwave signal generator using an AM modulator. Assuming that the lock-in frequency ($\omega_b$) is much smaller than the signal generator frequency ($\omega_c$), (i.e. $\omega_c >> \omega_b$) the output signal of AM modulator can be written as:

$$J_{c0} \sin(\omega_c t + \phi_1)\{1 + m \sin(\omega_b t)\} \qquad (1)$$

where $m$ is the AM modulation constant equals to the ratio of the lock-in to signal generator amplitude, and $J_{c0}$ is the current density. The frequency of the signal generated by the lock-in amplifier was set to 931.7 Hz and the signal generator frequency was swept from 10 MHz to 1 GHz in 0.5 MHz increments. When the frequency of the signal generator is close to the domain wall resonance frequency, the domain wall exhibits a resonant gyrotropic motion, and the resistance across the antivortex in the nanostructure would contain an oscillatory component due to the anisotropic magnetoresistance (AMR) effect [7]. The resultant changes in the resistance across the antivortex in the nanostructure could be written as:

$$R_0 \sin(\omega_c t + \phi_2)[J_{c0} \sin(\omega_c t + \phi_1)\{1 + m \sin(\omega_b t)\}]. \tag{2}$$

Here $R_0$ is the changes in the resistance due to the domain wall gyrotropic motion that could be only excited by the high frequency component of the input signal. The amplitude of the resultant sinusoidal voltage having a frequency of $\omega_b$, which is measured by the lock-in amplifier, is proportional to:

$$\frac{1}{2} m R_0 J_{c0} \cos(\phi_2 - \phi_1) \tag{3}$$

This is the phase difference between the input signal and the change in the device resistance.

In order to nucleate the domain wall, a perpendicular magnetic field ($z$-direction) was applied and the lock-in voltage was measured. As can be seen in the lock-in voltage in Fig. 1(b), above certain value of magnetic field ($H_c = 1275$ Oe), distinct peaks around 73.5 MHz appear at the output signal which are not observable below $H_c$. Subsequently, we performed magnetic force microscopy (MFM) on the sample in the area between the contacts and indeed we observed a domain wall as shown in Fig. 1(c). Therefore, by using this measurement setup, one can detect the magnetization dynamics and the threshold field for nucleation of the domain wall.

In order to see the effect of the amplitude of lock-in voltage on the output spectra of the domain wall, the lock-in voltage has been changed from 100 to 500 mV for a fixed signal generator voltage of 500 mV corresponding to $m = 0.2$ up to $m = 1$. As can be seen in Fig. 2(a), the output spectra of the domain wall dynamics is almost independent of lock-in voltages which is in agreement with our expectation that only the high frequency component of the input signal should excite the dynamics inside the domain wall. The main frequency component of the output voltage appears at 73.5 MHz, which is close to the resonance frequency of a vortex wall with a comparable size [19]. In addition, two more peaks are observable at 117.5 and 166.5 MHz, which could be associated with nonlinear eigenmodes of the domain wall. These results are consistent with nonlinear vortex wall dynamics that has been studied theoretically [20], and experimentally [21], in which the trajectory of the vortex deviates from a circular path to an elliptic path and multi-eigenfrequencies were reported.

Increasing the signal generator amplitude increases the input current density and it will ideally excite higher-order modes of the domain wall gyration. The result of domain wall response for different signal generator amplitudes with a fixed $m$ of 0.95 is shown in Fig. 2(b). As seen, for small signal generator amplitudes, the current density is too low to excite dynamics inside the domain wall. For a signal generator voltage of 500 mV, the current density in the branches of the nanostructure is approximately $3 \times 10^7$ A/cm$^2$. The second and third peaks above the main peak at 73.5 MHz are observable for signal generator voltages greater than 300 mV. Nonlinear dynamics of the antivortex is thus seen to be excited by the injection of high current density, as previously reported for the vortex wall [22].

Measurements of the resonance frequency of antivortices have been performed on nanostructures with three different sizes as shown in Fig. 2(c). Each ferromagnetic nanostructure was uniformly scaled, such that the dimension marked $D$ in Fig. 2(c) is 720 nm, 480 nm, and 410 nm. The main resonance frequencies were found to be 73.5, 112.3 and 176.4 MHz in the devices whose length along $D$ is 720 nm, 480 nm, and 410 nm, respectively [23]. Hence, a decrease in the device size results in an increase in the resonance frequency, which is consistent with results of the vortex wall [7].

The effect of the magnetic field on the dynamics of the domain wall has been also studied. The magnetic field was applied in the $x$-direction, and the gyration dynamics of the antivortex was measured by sweeping the frequency of the signal generator, delivering power at an amplitude of 500 mV as can be seen in Fig. 2(d). It is found that the output spectra of the domain wall do not change with the magnetic field up to 570 Oe. This results are in line with the previously reported results of vortex wall studies [19, 24]. Above 570 Oe, the frequency spectrum of the domain wall changes and the position of the main peak shifts from 73.6 to 247.5 MHz. Furthermore, the amount of changes ($\Delta V$) in the amplitude of output signal at the resonance frequency varies from around 11 to 13.6 µV. This could be due to the change in the domain wall structure in large magnetic fields (above 570 Oe). The $\Delta V$ is related to the domain wall gyrotropic amplitude and the AMR gradient across the domain wall [8], and thus a 23% variation in $\Delta V$ from 11 µV to 13.6 µV may be explained by the transformation of the domain wall configuration from an antivortex to other domain wall [25]. In order to better understand this phenomenon, the magnetic field was removed, and the domain wall frequency spectrum was measured again. It is found that the domain wall preserves its structure and the main mode at 247.5 MHz is still measureable [23]. Furthermore, even with the application of a magnetic field up to 2 kOe in the $x$-direction, significant changes in the domain wall resonance frequency was not observed. Therefore, we conclude that the new configuration of the domain wall is a stable state. Applying a magnetic field in the $x$-direction, the antivortex core is displaced in the $y$- direction, and if the external field is large enough, it is possible to annihilate the antivortex with the nucleation of a new domain wall [23, 26].

In a vortex or antivortex wall, the domain wall motion has a two dimensional trajectory. Depending on the characteristics of the injected current, the domain wall dynamics would either be under steady state precession in a two dimensional trajectory for a sinusoidal current excitation, or it can be a two dimensional damped trajectory for a pulsed current injection. [23] In the present experiments, the frequency of the sinusoidal input was increased by 0.5 MHz every 600 ms slow enough compared to the transient response of the domain wall to a sinusoidal current which is usually less than 100 ns, therefore, the effect of the domain wall transient response is negligible.

Similar to a vortex wall, the antivortex wall resonance frequency may be written as $f_{AV} = k_M/2\pi G_0$ [27], where $k_M$ is the antivortex stiffness due to displacement of the antivortex core from its equilibrium position and $G_0$ is the gyration vector amplitude. The gyration vector amplitude could be written as $G_0 = 2\pi qpLM_s/\gamma$, in which, $q = -1$ for antivortex, $p (= \pm 1)$ is the antivortex core polarity, and $L$ is the antivortex thickness [24]. The stiffness is related to the magnetic susceptibility $\chi_M = dM_x/dH_x$ by [24, 27]

$$k_M = \frac{\pi L M_s^2 \xi^2}{\chi_M} \qquad (4)$$

where $\xi$ is a parameter which describes the type of the boundary condition ($\xi \approx 1$) [24]. To calculate $\chi_M$, $1/\chi_M = 2\beta[\ln(8/\beta)-0.5]$ has been used [24], where $\beta (= L/R)$ is the aspect ratio of the antivortex thickness ($L$) over radius ($R$). By using the antivortex $D$ parameter as defined in Fig. 2 (c) instead of vortex diameter and the properties of Permalloy ($M_s = 8 \times 10^5$ A/m and $\gamma = 1.76 \times 10^2$ GHz/T), a resonance frequency of 83 MHz is calculated, which is in good agreement with the present experimental result of 73.6 MHz.

In order to study the domain wall transient response, we have used the circuit configuration shown in Fig. 3(a). A short voltage pulse was applied to displace the domain wall from its equilibrium position and excite antivortex dynamics. The antivortex dynamics is much slower than the excitation pulse width. Therefore, the antivortex gyrotropic motion remains after the excitation has removed, and it can be detected by applying a dc current through a bias tee in the excitation port. The voltage pulse has an 80 ps rise and fall time, and is negligible in comparison to the time response of antivortex dynamics, which is of the order of several tens of nanoseconds. The dc current (50 μA) with a current density of around $6 \times 10^6$ A/cm$^2$, is sufficiently small in comparison with the excitation current density and does not affect the domain wall dynamics. The output signal is measured by a Tektronix 6 GHz real time oscilloscope. The voltage pulses have a repetition of 1 kHz and the output signal is averaged 10000 times in the oscilloscope to improve the signal-to-noise ratio. Fig. 3(b) shows the output signal for an excitation pulse width of 5 ns, and amplitude of 4 V ($J$=1.2×10$^8$ A/cm$^2$). As can be seen, close to the excitation pulse, the output signal is complex and contains high frequency components, while the behavior of the output signal is similar to a damped sinusoidal away from the excitation. After removing the delay between the input pulse and output signal, the input excitation has been overlayed in Fig. 3(b) to show the origin of high-frequency components mostly in the first 10 ns of the transient response. At the rise and fall time of the input pulse, there is a sharp rise at the output signal follows by a high frequency component. This behavior is similar to the time resolve spin wave measurement by a pulse voltage [28] and could be associated to the spin wave excitation.

A curve fitting has been performed on the output signal to calculate the damping ratio and damped resonance frequency of the motion. The formula used in the curve fitting is $V = Ae^{-\Gamma t}\sin(2\pi ft + \Phi) + V_0$, with fitting parameters $A$ = 0.0158 volt, $\Gamma$ = 3.02×10$^7$ Hz, $f$ = 70.36

MHz, $\Phi = 0.539$ radian, and $V_0 \sim 0$ V. The damped resonance frequency was found to be 70.36 MHz is very close to the resonance frequency measured by the homodyne technique (73.5 MHz).

In conclusion, the dynamics of an antivortex wall has been studied experimentally using two different electrical excitation methods such as resonance sinusoidal and transient pulse excitations. In resonance excitation, by using the domain wall rectification effect, the eigenmodes of antivortex walls are measured. It is found that nonlinear modes of the antivortex could be excited for large excitation amplitudes. For the pulse excitation of the antivortex, the damped frequency of transient response is very close to the main eigenfrequency of the antivortex wall measured by resonance excitation. Furthermore, the transient response of the antivortex accompanies spin wave excitation inside the ferromagnetic structure. Our demonstration of all-electrical measurement paves the way toward a better understanding of antivortex dynamics in both frequency and time domains such as antivortex wall nucleation, the nonlinear behavior, and coexistence of the spin wave.

This work is supported by the Singapore National Research Foundation under CRP Award No. NRF-CRP 4-2008-06.

[*]Electronic address: eleyang@nus.edu.sg

References:

[1]     R. Hertel, S. Gliga, M. Fahnle, and C. M. Schneider, Phys. Rev. Lett. **98**, 117201 (2007).

[2]     A. Drews, B. Kruger, G. Meier, S. Bohlens, L. Bocklage, T. Matsuyama, and M. Bolte, Appl. Phys. Lett. **94**, 062504 (2009).

[3]     M. Bolte, G. Meier, B. Kruger, A. Drews, R. Eiselt, L. Bocklage, S. Bohlens, T. Tyliszczak, A. Vansteenkiste, B. Van Waeyenberge, K. W. Chou, A. Puzic, and H. Stoll, Phys. Rev. Lett. **100**, 176601 (2008).

[4]     J. Raabe, C. Quitmann, C. H. Back, F. Nolting, S. Johnson, and C. Buehler, Phys. Rev. Lett. **94**, 217204 (2005).

[5]     A. Bisig, J. Rhensius, M. Kammerer, M. Curcic, H. Stoll, G. Schutz, B. Van Waeyenberge, K. W. Chou, T. Tyliszczak, L. J. Heyderman, S. Krzyk, A. von Bieren, and M. Klaui, Appl. Phys. Lett. **96**, 152506 (2010).

[6]     R. Moriya, L. Thomas, M. Hayashi, Y. B. Bazaliy, C. Rettner, and S. S. P. Parkin, Nat. Phys. **4**, 368 (2008).

[7]     S. Kasai, Y. Nakatani, K. Kobayashi, H. Kohno, and T. Ono, Phys. Rev. Lett. **97**, 107204 (2006).

[8]     J. S. Kim, O. Boulle, S. Verstoep, L. Heyne, J. Rhensius, M. Kläuil, L. J. Heyderman, F. Kronast, R. Mattheis, C. Ulysse, and G. Faini, Phys. Rev. B **82**, 104427 (2010).

[9]     S. Gliga, R. Hertel, and C. M. Schneider, J. Appl. Phys. **103**, 112501 (2008).

[10]     X. J. Xing, Y. P. Yu, S. X. Wu, L. M. Xu, and S. W. Li, Appl. Phys. Lett. **93**, 202507 (2008).

[11]     S. Gliga, M. Yan, R. Hertel, and C. M. Schneider, Phys. Rev. B **77**, 060404 (2008).

[12]     J. Miguel, J. Sanchez-Barriga, D. Bayer, J. Kurde, B. Heitkamp, M. Piantek, F. Kronast, M. Aeschlimann, H. A. Durr, and W. Kuch, J. Phys.: Condens. Matter **21**, 496001 (2009).

[13]     K. Shigeto, T. Okuno, K. Mibu, T. Shinjo, and T. Ono, Appl. Phys. Lett. **80**, 4190 (2002).

[14]     P. E. Roy, J. H. Lee, T. Trypiniotis, D. Anderson, G. A. C. Jones, D. Tse, and C. H. W. Barnes, Phys. Rev. B **79**, 060407 (2009).

[15]     T. Kamionka, M. Martens, K. W. Chou, M. Curcic, A. Drews, G. Schutz, T. Tyliszczak, H. Stoll, B. Van Waeyenberge, and G. Meier, Phys. Rev. Lett. **105**, 137204 (2010).

[16]     K. Kuepper, M. Buess, J. Raabe, C. Quitmann, and J. Fassbender, Phys. Rev. Lett. **99**, 167202 (2007).

[17]     A. Yamaguchi, H. Miyajima, T. Ono, Y. Suzuki, and S. Yuasa, Appl. Phys. Lett. **91**, 132509 (2007).

[18]     A. Thiaville, and Y. Nakatani, J. Appl. Phys. **104**, 093701 (2008).

[19]     V. Novosad, F. Y. Fradin, P. E. Roy, K. S. Buchanan, K. Y. Guslienko, and S. D. Bader, Phys. Rev. B **72**, 024455 (2005).

[20]     K. Y. Guslienko, R. Rafael Hernández, and O. Chubykalo-Fesenko, Phys. Rev. B **82**, 014402 (2010).

[21]     X. M. Cheng, K. S. Buchanan, R. Divan, K. Y. Guslienko, and D. J. Keavney, Phys. Rev. B **79**, 172411 (2009).

[22]     D. Bedau, K, M, M. T. Hua, S. Krzyk, U. Rüdiger, G. Faini, and L. Vila, Phys. Rev. Lett. **101**, 256602 (2008).

[23]     See EPAPS Document for supplementary text and figures. For more information on EPAPS, see http://www.aip.org/pubservs/epaps.html.

[24]     K. Y. Guslienko, B. A. Ivanov, V. Novosad, Y. Otani, H. Shima, and K. Fukamichi, J. Appl. Phys. **91**, 8037 (2002).

[25]     M. Hayashi, L. Thomas, C. Rettner, R. Moriya, X. Jiang, and S. S. P. Parkin, Phys. Rev. Lett. **97**, 207205 (2006).

[26]     L. Bocklage, B. Krüger, R. Eiselt, M. Bolte, P. Fischer, and G. Meier, Phys. Rev. B **78**, 180405 (2008).

[27]     R. L. Compton, T. Y. Chen, and P. A. Crowell, Phys. Rev. B **81**, 144412 (2010).

[28]     M. Covington, T. M. Crawford, and G. J. Parker, Phys. Rev. Lett. **89**, 237202 (2002).

Figure Captions

FIG. 1.   (a) A SEM image of the ferromagnetic structure and a schematic representation of the electric circuit used for the measurement of the antivortex resonance frequency. (b) The frequency spectrum of the domain wall for different values of the perpendicular magnetic field with a 3 μV voltage offset for each data set. (c) MFM image of the antivortex wall.

FIG. 2.   (a) Frequency spectra of the antivortex wall for different lock-in amplifier output voltages normalized by $m$ with a 6 μV voltage offset for each data set. (b) The frequency spectrum of the antivortex for different values of the signal generator amplitudes in a logarithmic scale. (c) The resonance frequency of the antivortex for different device sizes. (d) The effect of the in-plane magnetic field in the $x$-direction on the frequency spectrum with a 12 μV voltage offset for each data set.

FIG. 3.   (a) The electric circuit configuration for the measurement of the transient response. (b) The measured output signal with the corresponding excitation pulse and the curve fitting data.

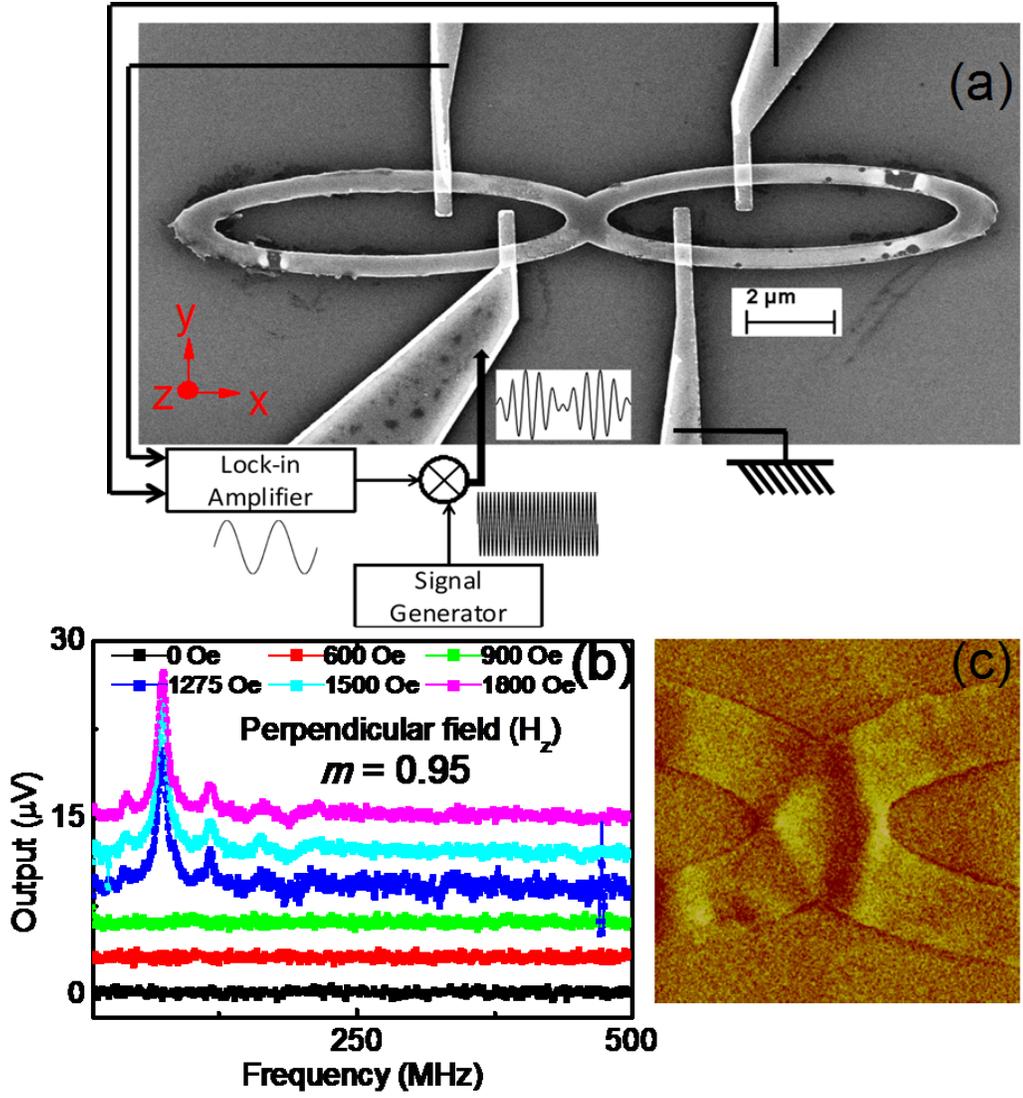

Figure 1.

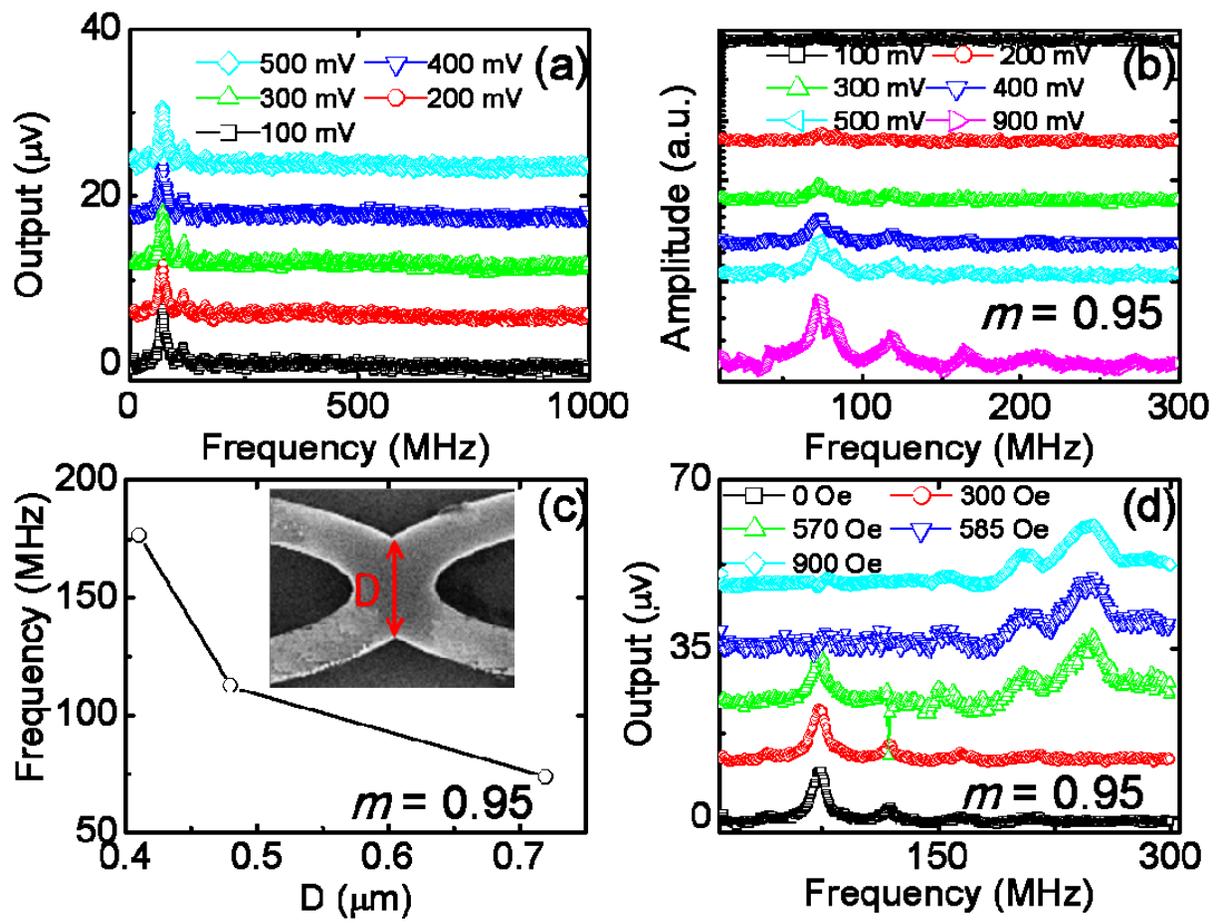

Figure 2.

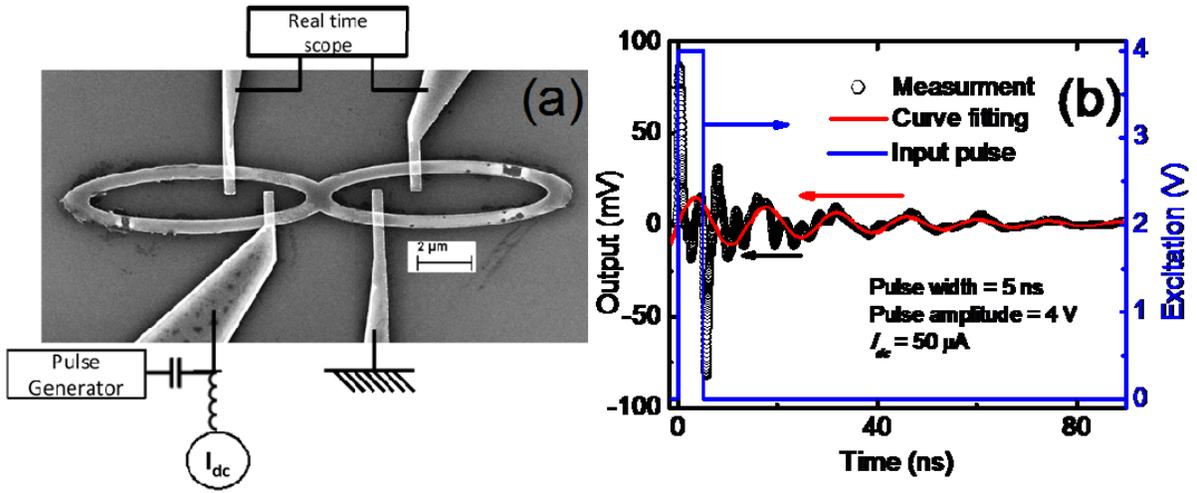

Figure 3.